\documentstyle[12pt]{article}
\advance\topmargin by -1.6cm

\newenvironment{eq}[1]
{\[\begin{array}{#1}}{\end{array}\]}

\let\rvec=\vec        

 \def\({\Bigl(}
\def\){\Bigr)}
 \def\|{\Big|}
\def\then{\Rightarrow}
 \def\o{\circ}

\def\x{\times}

\def\ox{\otimes}

\def\pl{{~\oplus~}}

\def\PL{\displaystyle \bigoplus}
\def\SUM{\displaystyle \sum}

\def\mid{\big\bracevert}

\def\sub{\subseteq}
\def\subnoteq{\subset}

\def\and{\wedge}

\def\AND{\displaystyle\bigwedge}

\def\OD{\displaystyle\bigvee}

\def\rin{{\,\in\kern-.42em\in}}
 
 \def\diag{{\,{\rm diag}\,}}

\def\spec{\,{\rm spec}\,}

\def\tr{{\,{\rm tr }\,}}
\def\det{\,{\rm det }\,}
\def\id{\,{\rm id}}

\def\sqsub{\sqsubseteq}
\def\sqsup{\sqsupseteq}

\def\full{ \rule{2.55mm}{2.55mm}  }

\def\empty{\Box}

\def\sx{~\rvec\x~\!}

\def\A{{\,{\rm A\kern-.55emA}}}
\def\B{{\,{\rm I\kern-.2emB}}}
\def\C{{\,{\rm I\kern-.55emC}}}
\def\E{{\,{\rm I\kern-.2emE}}}
\def\G{{\,{\rm I\kern-.55emG}}}
\def\H{{{\rm I\kern-.2emH}}}
\def\I{{\,{\rm I\kern-.2emI}}}
\def\K{{\,{\rm I\kern-.2emK}}}
\def\L{{\,{\rm I\kern-.2emL}}}
\def\M{{\,{\rm I\kern-.16emM}}}
\def\N{{\,{\rm I\kern-.16emN}}}
\def\Q{{\,{\rm I\kern-.5emQ}}}
\def\R{{{\rm I\kern-.2emR}}}
\def\S{{\,{\rm I\kern-.42emS}}}
\def\T{{\,{\rm I\kern-.37emT}}}
\def\UU{{\,{\rm I\kern-.51emU}}}
\def\Z{{\,{\rm Z\kern-.32emZ}}}

\def\al{\alpha}  \def\be{\beta} 
\def\de{\delta}  \def\ep{\epsilon}  \def\ze{\zeta}
    
\def\ka{\kappa}   \def\la{\lambda}   
\def\De{\Delta}    \def\Om{\Omega}
\def\phi{\varphi}
 \def\Ga{\Gamma}

\def\latt{\underline{\bf latt}}

\def\vec#1{\underline{\bf vec}_{#1}}

\def\GL{{\bf GL}}
\def\SL{{\bf SL}}
\def\U{{\bf U}}

\def\O{{\bf O}}
\def\SU{{\bf SU}}

\def\SO{{\bf SO}}

 \def\D{{\bl D}}
\def\AL{{\bf AL}}

\def\Ffo#1{\angle{#1}_{{}_{{\rm F}}}   }  
\def\d#1{{\check{#1}}}
\def\angle#1{\langle#1\rangle}

\def\rstate#1{|#1\rangle}
\def\lstate#1{\langle#1|}

\def\ro#1{{\rm #1}}
\def\bl#1{{\bf {#1}}}
\def\cl#1{{\cal #1}}
\def\ul#1{\underline{#1}}

\def\ol#1{\overline{#1}}

\def\dprod#1#2{\langle#1,#2\rangle}
\def\sprod#1#2{\langle#1|#2\rangle}

\def\com#1#2{\lbrack#1,#2\rbrack}

\def\acom#1#2{\{#1,#2\}}

\def\map{\longrightarrow}

\def\lrmap{\leftrightarrow}

\def\mape{\longmapsto}

\def\logic{{\ul{\bf logic}}}

\begin{document}

\begin{titlepage}
\hfill MPI-PhT/2002-16


\vskip25mm
\centerline{\bf PROBABILITY COLLECTIVES}

\centerline{\bf  FOR UNSTABLE PARTICLES}
\vskip1cm
\centerline{
Heinrich Saller\footnote{\scriptsize
hns@mppmu.mpg.de}
}
\centerline{Max-Planck-Institut f\"ur Physik}
\centerline{Werner-Heisenberg-Institut}
\centerline{M\"unchen}
\vskip25mm

\centerline{\bf Abstract}
\vskip5mm
Unstable particles, together with their stable decay products,
constitute probability collectives which are defined
as Hilbert spaces with  dimension higher than one, nondecomposable
in a particle basis.
Their structure is considered in the framework of Birkhoff-von Neumann's
Hilbert subspace lattices. Bases with  particle states
are related to bases with a diagonal scalar product
by a Hilbert-bein involving the characteristic decay parameters (in some
analogy to the $n$-bein structures of metrical manifolds).
Probability  predictions as expectation values,
involving unstable particles, have to take into account
all members of the higher dimensional collective.
E.g., the  unitarity structure  of the $S$-matrix for an unstable particle
collective
can be established  by a transformation with its Hilbert-bein.

\vskip1cm

\end{titlepage}

{\small \tableofcontents}

\newpage

\section{Stable Particle Hilbert Spaces}

The Hilbert space used for stable particles
with mass $m$,  momentum $\rvec p$ and
possible homogeneous  degrees of freedom $a=1,2,\dots, K$
- including  particles and antiparticles with spin and internal
degrees of freedom - comes with creation operators
$\ro u^a(m,\rvec p)$
and annihilation operators $\ro u_a^\star (m,\rvec p)$.
To have the involved concepts and notations at hand, it is shortly reviewed
by repeating its construction.

The underlying quantum structure for
Bose (commutator $\ep=-1$) and Fermi (anticommutator $\ep=+1$)
\begin{eq}{l}
\com{\ro u^\star}{\ro u}_\ep =1,~~
\com{\ro u}{\ro u}_\ep
=0=\com{\ro u^\star}{\ro u^\star}_\ep
\end{eq}comes with the time translation behavior of
the Bose and Fermi harmonic oscillator as implemented
by the Hamiltonian
\begin{eq}{l}
H= E{\com{\ro u}{\ro u^\star}_{-\ep}\over 2}
\end{eq}involving the quantum-opposite  commutator
and a frequency (energy)  scale $ E>0$.

Creation and annihilation operators build by the linear combinations of
their products the  quantum algebras $\bl Q_\ep(\C^2)$, countably infinite dimensional
for Bose and 4-dimensional for Fermi
\begin{eq}{l}
\bl Q_\ep(\C^2)=\C[\ro u,\ro u^\star]/
\hbox{ modulo }\left\{\begin{array}{l}
\com{\ro u^\star}{\ro u}_\ep-1\cr
\com{\ro u}{\ro u}_\ep,~~
\com{\ro u^\star}{\ro u^\star}_\ep\end{array}\right\}
\cong\left\{\begin{array}{lll}
\C^{\aleph_0}&\hbox{Bose},&\ep=-1\cr
\C^4&\hbox{Fermi},&\ep=+1\cr\end{array}\right.\cr
\hbox{basis of }
\bl Q_\ep(\C^2):~~
\left\{\begin{array}{ll}
\{\ro u^k(\ro u^\star)^l\mid k,l=0,1,\dots\},&\hbox{Bose}\cr
\{1,\ro u,\ro u^\star,\ro u\ro u^\star\},&\hbox{Fermi}\cr
\end{array}\right.\cr
\end{eq}The given bases contain eigenvectors of the time translations,
 ${d\over dt}a=[iH,a]$
\begin{eq}{l}
 [H,\ro u^k(\ro u^\star)^l]=(k-l) E \ro u^k(\ro u^\star)^l
\end{eq}

The Fock state $\Ffo{~~}$,
a conjugation compatible linear form  of the quantum algebras,
 is induced by the scalar product $\sprod {\ro u}{\ro u}=1$
 of  the 1-dimensional basic vector space $\C\ro u$.
The scalar product  invariance group $\U(1)$ contains  the irreducible
time translation representation $t\mape e^{it E}\in\U(1)$, generated by the
Hamiltonian $H$
\begin{eq}{l}
\bl Q_\ep(\C^2)\ni a\mape \Ffo a\in\C,~~\left\{\begin{array}{rl}
\Ffo{(\ro u^\star\ro u)^k}=(\Ffo{\ro u^\star\ro u})^k=1&k=0,1,\dots\cr
\Ffo{(\ro u^\star)^l\ro u^k}=0&\hbox{for }k\ne l\end{array}\right.
\end{eq}The Fock space $\hbox{Fock}_\ep(\C^2)$ is a quotient space of the
quantum algebra, constituted by the classes with
respect to the elements with trivial scalar product
(the annihilation left ideal in the quantum algebra)
\begin{eq}{l}
\{a\in\bl Q_\ep(\C^2)\mid\Ffo{aa^\star}=0\}=\bl Q_\ep(\C^2)\ro u^\star,~~
\hbox{Fock}_\ep(\C^2)=\bl Q_\ep(\C^2)/\bl Q_\ep(\C^2)\ro u^\star
\end{eq}$\hbox{Fock}_\ep(\C^2)$ has a definite scalar product.
The classes are called state vectors
$\rstate a$. The class $\rstate 0$ (zero quantum state vector)
of the algebra unit $1$ is
the harmonic oscillator ground state. It is a cyclic vector for the
quantum algebra action with the annihilation property
$\ro u^\star\rstate 0=0$.
The state vectors $\rstate k$ with  $k$ quanta constitute a
Fock space  basis, they are  time translation eigenvectors
\begin{eq}{l}
\hbox{ground state }\rstate 0=1+\bl Q_\ep(\C^2)\ro u^\star,~~
\rstate k={\ro u^k\over \sqrt{k!}}\rstate 0,~~
H\rstate k=(k-{\ep\over2}) E\rstate k\cr

\hbox{basis of $\hbox{Fock}_\ep(\C^2)$: }\left\{ \begin{array}{ll}
\{\rstate k\mid k=0,1,\dots\}&\hbox{Bose}\cr
\{\rstate 0, \rstate 1\}&\hbox{Fermi}\cr\end{array}\right\}
,~~\sprod kl=\de_{kl}\cr
\end{eq}

The Bose structure in quantum mechanics - not the Fermi structure -
 allows a position-momentum interpretation
\begin{eq}{l}
\hbox{Bose only: } \bl x={\ro u^\star+\ro u\over \sqrt2},~~
 i\bl p={\ro u^\star-\ro u\over\sqrt 2}\then\left\{\begin{array}{l}
[\ro u^\star,\ro u]=1=[i\bl p,\bl x]\cr
H= E{\acom{\ro u}{\ro u^\star}\over 2}= E{\bl p^2+\bl x^2\over
2}\end{array}\right.
\end{eq}The familiar Schr\"odinger wave functions
$\rstate k\cong\psi_k(x)$ are orbits of
the position translation $\R=\spec\bl  x\ni x\mape\psi(x)\in\C$.

The Fermi Fock space  is a Hilbert space
as well as the completion of
the Bose Fock space.
The Fock vector spaces
 are the direct sum of the
  totally symmetric and antisymmetric tensor powers
for Bose and Fermi resp. of
the complex 1-dimensional Hilbert space
$\C\rstate 1$ with the 1-quantum  state vectors of  energy  $E$
denoted by $\rstate1=\rstate E$
\begin{eq}{l}
\begin{array}{rll}
\hbox{Fock}_-(\C^2)&={\OD}W_-( E)&\cong\C^{\aleph_0}\cr
\hbox{Fock}_+(\C^2)&={\AND}W_+( E)&\cong\C^2\end{array}\hbox{with}
\left\{\begin{array}{rl}
W_\ep( E)&=\C\rstate{ E}\cong\C\cr
\rstate  E&=
\ro u( E)\rstate 0,~~\sprod  E E=1\cr
\end{array}\right.
\end{eq}Such Fock spaces will be used for different
energies (frequencies)
$ E=p_0=\sqrt{m^2+\rvec p^2}$.

For a stable particle with mass $m$, momentum
$\rvec p$ and homogeneous  degrees of freedom
$a=1,\dots,K$ one works with
a direct sum-integral of Hilbert spaces, integrating with
a Lorentz invariant measure  the 1-quanta Hilbert spaces
for all momenta
\begin{eq}{rl}
{\PL_{a=1}^K}\int_{\R^3} {d^3p\over(2\pi)^3p_0} W_\ep^a(m,\rvec p):&
[\ro u_b^\star(m,\rvec q),\ro u^a(m,\rvec p)]_\ep
=\de^a_b(2\pi)^3p_0\de(\rvec p-\rvec q)\cr
&\hbox{with }p_0=\sqrt{m^2+\rvec p^2}\cr
&W^a_\ep(m,\rvec p)=\C\rstate {m,\rvec p,a},~~\rstate {m,\rvec p,a}=
\ro u^a(m,\rvec p)\rstate 0\cr
&
\sprod{m,\rvec q,b}{m,\rvec p,a}
=\de^a_b(2\pi)^3p_0\de(\rvec p-\rvec q)\cr
\end{eq}

Up to the overcountably infinite
dimensional momentum dependence $\C^{\R^3}$
the 1-quantum basic Hilbert spaces used are a direct sum of a Bose and a Fermi
space with - for stable particles - orthogonal subspaces for
 different masses
 \begin{eq}{rl}
W=W_+\pl W_-:&
W_\ep={\PL_{A=1}^{s}}W_\ep(m_A),~~
W_\ep(m)={\PL_{a=1}^K}W_\ep^a(m)\cong\C^K\cr
&\sprod {m_B,b}{m_A,a}=\de_{AB}\de^a_b
\end{eq}The corresponding multiquanta states -
generalizing the  state vectors $\rstate k\in\hbox{Fock}_\ep(\C^2)$
above - are appropriately defined tensor
products.

\section{Unstable States and Particles (Part 1)}

To introduce into the later, more abstract sections, the
kaon system with the short and long lived
unstable neutral kaon  is given as an illustration.

\subsection{The Collective of  Neutral Kaons}

The system of the two neutral $K$-mesons $K^0  _{S,L}$
(short and long)
with the mass denoted state vectors  $\rstate {M  _{S,L}}$
\begin{eq}{l}
M=M^0+i{\Ga\over 2},~~\Ga>0
\end{eq}spans a 2-dimensional Hilbert space.
The kaon particles
are no  CP-eigenstates  $\rstate{K_{\pm}}$ to which they can be transformed by
an invertible $(2\x 2)$-matrix, called the Hilbert-bein of the neutral
kaon-system
\begin{eq}{rl}
{\scriptsize\pmatrix{\rstate {M_S  }\cr\rstate {M_L  }\cr}}
&=\xi_2{\scriptsize\pmatrix{
\rstate {K_+}\cr\rstate {K_-}\cr}},~~\xi_2\in\GL(\C^2)
\end{eq}Under the assumption of CPT-invariance
the matrix is symmetric and parame\-tri\-zable by
two complex numbers
including a normalization factor
$N$
\begin{eq}{l}
\xi_2={1\over N\sqrt{1+|\ep|^2}}
{\scriptsize\pmatrix{1&\ep\cr \ep&1\cr}}
,~~\ep,N\in\C\cr
\end{eq}There are no
observable particles connected with the CP-eigenstates.

The time development is implemented by a Hamiltonian, non-hermitian for
unstable particles $H_2\ne H_2^\star$
\begin{eq}{l}
\hbox{for }t\ge0:~
{d\over dt}
{\scriptsize\pmatrix{\rstate {K_+}\cr\rstate {K_-}\cr}}=iH_2
{\scriptsize\pmatrix{\rstate {K_+}\cr\rstate {K_-}\cr}},~~
{d\over dt}{\scriptsize\pmatrix{
\rstate {M_S  }\cr\rstate {M_L  }\cr}}=i\diag H_2
{\scriptsize\pmatrix{\rstate {M_S  }\cr\rstate {M_L  }\cr}}\cr
\end{eq}with the diagonal form for the energy eigenstates
\begin{eq}{l}
\xi_2H_2\xi_2^{-1}=\diag H_2=
{\scriptsize\pmatrix{M_S&0\cr 0&M_L\cr}}
\end{eq}

The CP-eigenstates  constitute an orthonormal basis
\begin{eq}{l}
\hbox{CP-eigenstates: }
{\scriptsize\pmatrix{
\sprod{K_+}{K_+}&\sprod{K_+}{K_-}\cr
\sprod{K_-}{K_+}&\sprod{K_-}{K_-}\cr}}
={\scriptsize\pmatrix{1&0\cr0&1\cr}}\cr
\end{eq}whereas the scalar product  of the energy eigenstates
is given by the  absolute square of the Hilbert-bein
\begin{eq}{rl}
\hbox{particles: }
\ze_2=\xi_2\xi_2^\star&={\scriptsize\pmatrix{
\sprod{M_S  }{M_S  }&\sprod{M_S  }{M_L  }\cr
\sprod{M_L  }{M_S  }&\sprod{M_L  }{M_L  }\cr}}
=
{1\over |N|^2}
{\scriptsize\pmatrix{1&\de\cr
\de &1\cr}}\cr
\hbox{with }\de&={\ep+\ol\ep\over 1+|\ep|^2}, ~0\le|\de|\le 1
\end{eq}The experiments give a nontrivial
transition between the short and long lived kaon proportional to
 the real part of $\ep$.
Therefore  $\ze_2$  is not diagonal and  $\xi_2$ not  unitary
\begin{eq}{l}
 \de\sim 0.327\x10^{-2}\then \xi_2\notin\U(2)
\end{eq}

A decomposition of the unit operator in the 2-dimensional Hilbert space
can be written with
orthonormal bases, e.g. with  CP-eigenstates
\begin{eq}{l}
\bl 1_2
=\rstate{K_+}\lstate {K_+}+
\rstate{K_-}\lstate {K_-}\cr
\end{eq}or with the non-orthogonal particle  basis
which displays  the inverse scalar product matrix
\begin{eq}{rl}
\ze_2^{-1}&=
{|N|^2\over  1-\de^2}{\scriptsize\pmatrix{1&-\de\cr -\de&1\cr}}\cr
\then\bl 1_2&={|N|^2\over  1-\de^2}
\Bigl[\rstate  {M_S  } \lstate{M_S  }-\de \rstate {M_S  }\lstate{M_L  }
-\de \rstate {M_L  }\lstate{M_S  }+\rstate  {M_L  }\lstate{M_L  }\Bigr]\cr
\end{eq}

\subsection{Decay Collectives}

The two translation eigenstates (particles)
for  unstable kaons $\rstate{M_{S,L}}$ come together with their  decay products,
e.g.
$\rstate{\pi,\pi}$, $\rstate{\pi,\pi,\pi}$,
$\rstate{\pi,l,\nu_l}$, approximated as stable in the following.
All those particles together constitute an example for a  decay collective,
consisting of unstable decaying particles and their stable decay products.

In general, $d$ unstable states (particles)
$\{\rstate{M_\ka}\mid\ka=1,\dots,d\}$
spanning the space $\rstate {M}\cong\C^d$ with complex masses $M=M^0+i{\Ga\over 2}$,
$\Ga>0$, are considered together
with their stable decay modes, given by $s$ states
 (particles)
 $\{\rstate{m_a}\mid a=1,\dots,s\}$
  with real masses $m$ which span the space $\rstate {m}\cong\C^s$.
All those particles are assumed to span a  Hilbert space $W$
with dimension $n=d+s$.
Therein, the subspace $\rstate m$
has an orthonormal particle basis
\begin{eq}{l}
\sprod mm=\bl 1_s
\end{eq}There are orthonormal bases $\rstate U$ for the
$d$-dimensional complementary space $\rstate m^\bot\cong\C^d$
\begin{eq}{l}
{\scriptsize\pmatrix{
\sprod UU&\sprod Um\cr
\sprod mU&\sprod mm\cr}}=
{\scriptsize\pmatrix{
\bl 1_d&0\cr 0&\bl 1_s\cr}}
\end{eq}

The time development in the orthonormal basis
$\{\rstate U,\rstate m\}$ has the typical triangular
form with the diagonal
time development for
the stable particles
$m=m^\star=\diag m$ and a nondiagonal $(d\x s)$-matrix $D$ parametrizing
the decay structure
\begin{eq}{l}
\hbox{for }t\ge0:~~
{d\over dt}
{\scriptsize\pmatrix{\rstate {U}\cr\rstate {m}\cr}}=
iH_W
{\scriptsize\pmatrix{\rstate {U}\cr\rstate {m}\cr}},~~
H_W={\scriptsize\pmatrix{{H_d}&D\cr 0&m\cr}},~~{d\over dt}
\rstate {m}= m\rstate {m}
\end{eq}

The Hamiltonian cannot be hermitian, i.e. $H_W\ne H_W^\star$ -
otherwise all eigenvalues would be real. It can even not be
normal in the orthonormal basis, i.e. $H_WH_W^\star\ne H_W^\star H_W$ - otherwise it could be unitarily
diagonalized $\diag H_W=\xi H_W \xi ^{-1}$ with $\xi \in\U(n)$ and, therewith,
the energy eigenstates
 were necessarily  orthogonal with the scalar product matrix
$\xi \xi ^\star =\bl 1_n$.
The Hamiltonian $H_W$
 has to be diagonalizable, i.e. its minimal
polynomial has to have only order one zeros.
The nonunitary diagonalization matrix
\begin{eq}{l}
\xi_W H_W\xi_W ^{-1}=\diag H_W =
{\scriptsize\pmatrix{M&0\cr 0&m\cr}} ,~~\xi_W \not\in\U(n),~~M\ne M^\star
\end{eq}called the Hilbert-bein of
the decay collective,
is the product of a $(d\x d)$-matrix $\xi_d$ diagonalizing $H_d$
-  as exemplified in the kaon system of the former subsection -
and a triangular matrix
\begin{eq}{rl}
\xi_dH_d\xi_d^{-1}&= M ,~~
\xi_W =
{\scriptsize\pmatrix{\bl 1_d&w\cr 0&\bl 1_s\cr}}
{\scriptsize\pmatrix{\xi_d&0\cr 0&\bl1_s\cr}}
={\scriptsize\pmatrix{\xi_d&w\cr 0&\bl1_s\cr}}\cr\cr
\then
H_W&=\xi_W ^{-1}(\diag H_W) \xi_W =
{\scriptsize\pmatrix{H_d&\xi_d^{-1}(Mw-wm)\cr 0&m\cr}}\cr
\hbox{i.e. }
D&=\xi_d^{-1}(Mw-wm)
\end{eq}

The decaying particles
$\rstate {{M}}$
have projections both on the orthogonal states $\rstate U$ and on the
stable particles $\rstate m$
\begin{eq}{l}
\hbox{particles: }
{\scriptsize\pmatrix{\rstate {{M}}\cr\rstate {m}\cr}}=
{\scriptsize\pmatrix{\xi_d&w\cr 0&\bl1_s\cr}}
{\scriptsize\pmatrix{\rstate {U}\cr\rstate {m}\cr}}
={\scriptsize\pmatrix
{\xi_d\rstate {U}+w\rstate {m}\cr\rstate {m}\cr}}\cr
\end{eq}The scalar product matrix for the decay collective with the
$n=d+s$ particles
 arises from the diagonal matrix with the orthonormal states
and the decay channels
\begin{eq}{rl}
\ze_W=\xi_W \xi_W ^\star&= {\scriptsize\pmatrix{
\sprod{{M}}{{M}}&\sprod{{M}}{m}\cr
\sprod{ m}{{M}}&\sprod{ m}{m}\cr}}
=
{\scriptsize\pmatrix{\bl 1_d&w\cr 0&\bl 1_s\cr}}
{\scriptsize\pmatrix{\ze_d&0\cr 0&\bl 1_s\cr}}
{\scriptsize\pmatrix{\bl 1_d&0\cr w^\star&\bl 1_s\cr}}
=
{\scriptsize\pmatrix{
\ze_d +ww^\star & w\cr w^\star& \bl1_s\cr}}\cr
\hbox{ with }\ze_d&=\xi_d\xi_d^\star
\end{eq}The stable particles remain an orthogonal basis of the
subspace $\rstate m$.

The decomposition of the Hilbert space unit operator
in the non-orthogonal particle basis displays  the
inverse scalar product matrix
\begin{eq}{rl}
\ze_W^{-1}&=
{\scriptsize\pmatrix{\bl 1_d&0\cr -w^\star&\bl 1_s\cr}}
{\scriptsize\pmatrix{\ze_d^{-1}&0\cr 0&\bl 1_s\cr}}
{\scriptsize\pmatrix{\bl 1_d&-w\cr0&\bl 1_s\cr}}
={\scriptsize\pmatrix{
\ze_d^{-1}&-\ze_d^{-1}w\cr
-w^\star \ze_d^{-1}&\bl1_s+w^\star \ze_d^{-1} w\cr}}\cr\cr
\then\bl 1_n&=\rstate U\lstate U+\rstate m\lstate m\cr
&=\rstate M \ze_d^{-1}\lstate M
-\rstate M \ze_d^{-1} w\lstate m
-\rstate m w^\star \ze_d^{-1}\lstate M
+\rstate m (\bl1_s+w^\star \ze_d^{-1} w)\lstate m
\end{eq}

To define probabilities
 and expectation values for unstable particles,
a more general orientation with respect to the Hilbert space structures
involved will be useful.

\section
[Logic of Quantum Theory (A Short Review)]
{Logic of Quantum Theory\\(A Short Review)}

With Boole, Leibniz's dream of a formalization of logic which allows
to draw  conclusions in a
mechanical way - like arithmetic computation -
started to become realized.
Apparently, logic condenses the structures of our experiences
and, therewith, shows a close relationship to the formulations of physics.
With the paramount importance of the
complex linear superposition structure of quantum
theory the classical Boolean logic,
appropriate for classical phase space physics,
gave way to a quantum logic as formulated by Birkhoff and von Neumann.
As a consequence, the  probability structure, already
arising in classical physics, e.g. in thermostatistics,
 is not primary in quantum theory -
it comes, so to say,
as a square of a linear complex probability amplitude structure.

Nothing is new in the following section - it should serve as a short reminder
 and should introduce the concepts and notations used later on.
  In addition to Birkhoff-von Neumann's original article there is
Varadarajan's detailed textbook  which
can be consulted for a deeper information.

\subsection{Logics}
 The propositions
 of a logic are formalized as the elements of a lattice, i.e. of a set
with two  associative and commutative
inner compositions
(meet $\sqcap$ and {join} $\sqcup$)
which have an  absorptive relationship to each other
\begin{eq}{rl}
(L,\sqcap,\sqcup)\in\latt:&
a\sqcup(a\sqcap b)=a=a\sqcap(a\sqcup b)\hbox{ (absorptive) }\cr
\end{eq}Each lattice  carries its
 natural  order $a\sqsub b\iff a\sqcap b=a$.

A lattice with an  origin $\Box$ - it is unique
\begin{eq}{l}
\Box\sqsub a,\hbox{ i.e. }\Box=\Box\sqcap a\hbox{ for all }a\in L
\end{eq}allows the definition of
disjoint elements by $a\sqcap b=\empty$.

A complementary lattice has an involutive
contra-morphism relating meet and join with the origin as
meet for each lattice element and its complement
\begin{eq}{l}
L\map L,~~a\mape a^c,~~a^{cc}=a, ~\left\{\begin{array}{rl}
(a\sqcup b)^c&= a^c\sqcap b^c\cr
a\sqcap a^c&=\Box\hbox{ for all }a\in L\cr\end{array}\right.
\end{eq}The complement of the
origin is the  unique  end $\full$
\begin{eq}{l}
 \Box^c=\full\sqsup a,~~
a\sqcup a^c=\full\hbox{ for all }a\in L
\end{eq}

With an appropriate language for the logical concepts
a complemented lattice is used as a  logic
\begin{eq}{l}
(L,\sqcap,\sqcup,\Box,c)\in\logic:~~\left\{
\begin{array}{rl}
a\in L:&\hbox{proposition}\cr
\sqcap:&\hbox{conjunction (and, et)}\cr
\sqcup:&\hbox{adjunction (or, aut)}\cr
\sqsub:&\hbox{implication (then, ergo)} \cr
\empty:&\hbox{absurd proposition (falsehood, falsum)}\cr
a^c\in L:&\hbox{negation (not, non)}\cr
\full :&\hbox{self-evident proposition (truth, verum)}\cr\end{array}\right.\cr
\end{eq}

A lattice is {distributive} for
\begin{eq}{l}
a\sqcup(b\sqcap c)= (a\sqcup b)\sqcap (a\sqcup c)\cr
a\sqcap(b\sqcup c)= (a\sqcap b)\sqcup (a\sqcap c)\cr
\end{eq}Weaker than distributivity is {modularity},
a partial associativity for meet and join
\begin{eq}{l}
a\sqsub c\then a\sqcup(b\sqcap c)=(a\sqcup b)\sqcap c
\end{eq}

\subsection{Boolean Logics}

A distributive logic is, by Stone's theorem \cite{STO},
 isomorphic to a lattice of subsets $\cl M\sub 2^M=\{X\sub M\}$
with $2^M$ the power set of a set $M$. The lattice operations are
intersection and union, the negation uses the set complement
\begin{eq}{l}
\hbox{(distributive) }\logic\ni (L,\sqcap,\sqcup,\Box,c)
\cong (\cl M,\cap,\cup,\emptyset,C_M)
\end{eq}

The valuation of a Boolean logic employs probability measures,
i.e. disjoint additive mappings on the lattice with positive values
between $0$ for the falsum and $1$ for the verum
\begin{eq}{l}
 \mu:\cl M\map\R_+,~~\left\{\begin{array}{rl}
 \mu(\emptyset)&=0\cr
\mu(X\cup Y)&=\mu(X)+\mu(Y)\hbox{ for }X\cap Y=\emptyset\cr
\mu(M)&=1\end{array}\right.
\end{eq}

In the classical formulation of physics,
the propositions in the  corresponding Boolean logic
are subsets of the phase space of
a physical system. In deterministic classical mechanics, the
measurements are formalized by the numerical values of phase space
functions, i.e. the probabilities used are yes-no probabilities
on point subsets $\{(x,p)\}$ of the phase space
\begin{eq}{l}
X=\{(x,p)\}:~~\mu_X:\{\emptyset,X\}\map\{0,1\},~~\mu_X(X)=1
\end{eq}In thermostatistics coarser  subset lattices are used.

\subsection{Birkhoff-von Neumann Logics}

The subspaces $2^{\ol V}=\{W\sub V\mid\hbox{ closed subspace}\}$
 of a Hilbert space $V$ -
in the following only over the abelian fields $\K=\R,\C$ -
 with, again, the intersection as meet, but
 the span as join,
 the trivial space for the logical `falsum'
and the orthocomplementation for the  negation
constitute a linear logic
\begin{eq}{l}
(L,\sqcap,\sqcup,\Box,c)\cong (2^{\ol V},\cap,+,\{0\},\bot)
\in\logic\hbox{ (linear)}
\end{eq}A Hilbert state space of quantum mechanics is used for
a Birkhoff-von Neumann logic. It
extends the set union for a Boolean logic by
 the quantum characteristic linear superposition.
In the following, the relevant features of subspace lattices
are reviewed.

For dimension $n\ge2$ (where the vector space endomorphisms are
nonabelian) linear  lattices are not distributive (basis $\{e^i\}$)
\begin{eq}{rl}
W_i=\K e^i\cong\K,~~i=1,2,&\hbox{full space: }V=W_1+ W_2\cong\K^2\cr
&\hbox{diagonal space: }\De=\K (e^1+e^2)\cong\K\cr
(W_1+W_2)\cap\De=\De\ne(W_1\cap\De)&\hskip-2mm+(W_2\cap\De)=\{0\}+\{0\}
=\{0\}=W_1\cap W_2
\end{eq}A lattice with  finite dimensional subspaces is modular.

Linear lattices can
be `operationalized', i.e. they can be embedded
into the  endomorphisms of the full vector space $V$,
denoted by $\AL(V)$ -
a unital $\K$-algebra, for finite dimensions  $\AL(V)\cong V\ox V^T$ with the
dual space $V^T$.
Any idempotent  $\cl P$ (projector for $\cl P\ne 0$)
defines a subspace $W$ and
- by its kernel - a direct complement $W'$
\begin{eq}{l}
\AL(V)\ni \cl P=\cl P^2\mape
W=\cl P(V)\in 2^V\cr
V=W\pl W'\hbox{ with }W'= \cl P^{-1}(0)\in 2^V
\end{eq}One subspace can be defined by different projectors and
can have different complements - in the example above
with two different dual bases
\begin{eq}{rl}
\id_V&=\cl P_1+\cl P_2=e^1\ox\d e_1+e^2\ox\d e_2\cr
&=\cl P_1'+\cl P_\De=e^1\ox(\d e_1-\d e_2)+(e^1+e^2)\ox\d e_2\cr
W_1&=\cl P_1(V)=\cl P_1'(V),~~V=W_1\pl W_2= W_1\pl\De\cr
\end{eq}Uniqueness can be  obtained with  a dual isomorphism
$V\stackrel\ze\lrmap V^T$.

With a nondegenerate inner product
(symmetric bilinear or sesquilinear form)
\begin{eq}{rl}
\sprod{~}{~}:V\x V\map\K,&\ze(v,w)=\sprod vw=\ol{\sprod vw}\cr
&\sprod v{w+u}=\sprod vw+\sprod vu,~~\sprod v{\al w}=\al\sprod vw\cr
&\sprod vV=\{0\}\iff v=0\cr
\end{eq}each subspace has a unique
orthogonal subspace partner
\begin{eq}{l}
\bot:2^V \map 2^V,~~W\mape W^\bot=\{v\in V\mid \sprod Wv=\{0\}\}\cr
\end{eq}In a finite dimensional space $V$
orthogonality
defines an involution
\begin{eq}{l}
V\cong\K^n:~~W=W^{\bot\bot}
\end{eq}With a nondegenerate square,
projectors are bijectiveley
related to subspaces
\begin{eq}{l}
2^V\ni W\stackrel\ze\lrmap \cl P_W\in   \AL(V),~~W=\cl P_W(V)
\end{eq}i.e. the lattice of vector subspaces
can be considered to be operators.

The dual isomorphism allows the  bra-ket notation
(next subsection) wherewith
the projector  for a finite dimensional subspace $W\cong\K^k$
can be written with a
$W$-basis $\{e^\ka\}_{\ka=1}^k$ (summation convention)
\begin{eq}{rl}
\cl P_W&=
 \rstate{e^\ka}\ze_{ \la\ka}\lstate{e^\la}\hbox{ with }
\sprod{e^\la}{e^\mu}=\ze^{\mu\la},~~
\ze^{\mu\la}\ze_{\la\ka}=\de_\ka^\mu\cr
\tr_{\hskip-1mm V}~ \cl P_W&= \ze_{\la\ka}\ze^{\ka\la}=\de_\ka^\ka=d(W)\cr

\end{eq}

The involution defined by orthogonality is
not  complementary for an indefinite
nondegenerate square. E.g., in a 2-dimensional $\O(1,1)$-Minkowski
space with
Lorentz metric ${\scriptsize\pmatrix{1&0\cr0&-1\cr}}$
in a basis $\{e^0,e^3\}$,
time and position translations
$\T$ and $\S$ resp. are orthogonal to each other
whereas
the isotropic lightlike  subspaces $\L_\pm$ are self-orthogonal
\begin{eq}{rl}
\T^\bot&=(\R e ^0)^\bot=\R e^3=\S\cr
\al\be&\ne0\then
\R (\al e ^0+\be e^3)^\bot =\R ({1\over\al} e ^0+{1\over \be} e^3)
\then\L_\pm=\R (e ^0\pm e^3)= \L_\pm^\bot\cr
\cl P_{\T}&=\rstate{e^0}\lstate {e^0},~~
\cl P_{\S}=-\rstate{e^3}\lstate {e^3},~~
\cl P_{\L_\pm}={1\over 2}\rstate{e^0\pm e^3}\lstate {e^0\mp e^3}

\end{eq}A {complementary} linear lattice
 has to come with a definite square, i.e. with a scalar product,
 which is nondegenerate in each subspace
\begin{eq}{l}
\sprod vv=0\iff v=0\then \K^n\cong V= W+W^\bot=W\pl W^\bot=W\bot W^\bot
\end{eq}An inner product of $V\cong\C^n$ is
positive if, and only if, it is a product $\ze_n=SS^\star$
of an endomorphism $S\in\AL(V)$ and its $\U(n)$-hermitian $S^\star$.

Now the probability valuation of vector space sublattices:
With a scalar product  $\ze$, each nontrivial finite dimensional
vector space $W$  carries
 a {yes-no probability}  with a normalized  discriminant
\begin{eq}{rl}
\ze\succeq 0:&\mu_W:\{\emptyset,W\}\map \{0,1\}\cr
&\mu_W(W)=\det\ze_W=\det\sprod{e^\la}{e^\mu}=1\cr
\end{eq}The classical measure comes
as the positive scalar product.

The  Schr\"odinger wave functions - possible for Bose structures, e.g.
the harmonic Bose oscillator above $\rstate k\cong\psi_k(x)$ -
as position translation orbits
allow a `smearing out' of the probabilities
for the 1-dimensional subspaces (Hilbert rays $W=\C\rstate k$)
\begin{eq}{l}
\det\ze_W=
\sprod kk=\int_\R dx|\psi_k(x)|^2=1
\end{eq}to   position densities for the probability, here  $|\psi_k(x)|^2$.

For finite dimension $V\cong\K^n$, the trace
is an invariant linear
form with the trace of a projector giving the dimension
of the defined subspace
  \begin{eq}{rl}
\tr_{\hskip-1mm V}~:  \AL(V)\map\K,~~f\mape&\tr_{\hskip-1mm V}~ f\cr
&\tr_{\hskip-1mm V}~ \cl P=\dim_\K\cl P(V)\cr
\end{eq}The {expectation values $\cl E_W$ in
the subspace $W$} of the operating algebra elements
uses the trace, normalized with the dimension $\tr\cl P_W=d(W)$
\begin{eq}{rl}
\AL(V)\ni f\mape \cl E_W(f)&={1\over d(W)}\tr_{\hskip-1mm V}~\cl P_W\o f
={\ze_{\la\ka}\lstate{e^\la}f\rstate{e^\ka}\over d(W)}\in\K\cr
\end{eq}The expectation values in  $W$ of the
other subspaces use their operational form as  projectors
\begin{eq}{l}
\cl E_W: 2^{V}\map\R_+,~~\left\{\begin{array}{rl}
\cl E_W(U)=\cl E_W(\cl P_U)&= {1\over d(W)}\tr_{\hskip-1mm V}~\cl P_W\o \cl P _U\in
[0,d(U)]\cr
 \cl E_W(U_1\bot U_2)&=\cl E_W(U_1)+ \cl E_W(U_2)\cr
 U\sub W&\then  \cl E_W(U)={d(U)\over d(W)}\cr
\cl E_W(V)=\cl E_W(\id_V)&={1\over d(W)}\tr_{\hskip-1mm V}~\cl P_W
=1=\cl E_W(W)
\end{array}\right.
\end{eq}A familiar special case are the
 symmetric transition probabilities
between two  1-dimensional spaces ($i=1,2$)
\begin{eq}{l}
W_i=\K\rstate{e^i},~~
\cl P_{W_i}=\rstate{e^i}\lstate {e^i}
\then
\cl E_{W_1}(W_2)=\cl E_{W_2}(W_1)=|\sprod{e^1}{e^2}|^2\in[0,1]
\end{eq}

A basis formulation reads as follows
- each subspace comes with a basis (different indices)
\begin{eq}{l}
\left.\begin{array}{rl}
\cl P_W&= \rstate{e^\ka}\ze_{\la\ka}\lstate{e^\la}\cr
\cl P_U&= \rstate{e^A}\ze_{BA}\lstate{e^B}
\end{array}\right\}\then
\left\{\begin{array}{rl}
\cl P_W\o\cl P_U&=
 \rstate{e^\ka}\ze_{\la\ka}\sprod{e^\la}{e^A}\ze_{BA}\lstate{e^B}\cr
\tr_{\hskip-1mm V}~ \cl P_W\o\cl P_U&=
 \ze_{\la\ka}\ze_{BA}\sprod{e^\la}{e^A}\ol{\sprod{e^\ka}{e^B}}\cr
\end{array}\right.\end{eq}especially simple for Euclidean bases
\begin{eq}{rl}
\ze_{\la\ka}=\de_{\la\ka},~~
 \ze_{BA}=\de_{BA}
 \then \tr_{\hskip-1mm V} \cl P_W\o\cl P_U
  ={\SUM_{\la,A}} |\sprod{e^\la}{e^A}|^2\le d(W)d(U)
  \end{eq}

\subsection{Hilbert-Beins for Particle Collectives}

For a   basis $\{e^\ka\mid\ka=1,\dots,n\}$
of the Hilbert space $W\cong\C^n$ the scalar product gives the matrix
\begin{eq}{l}
\ze:W\x W\map\C,~~\sprod{e^\la}{e^\ka}=\ze^{\ka\la}=\ol{\ze^{\la\ka}}
\end{eq}with the inverse scalar product on the dual space
(linear forms) with the dual basis $\{e_\ka\mid\ka=1,\dots,n\}$
\begin{eq}{l}
\ze^{-1}:W^T\x W^T\map\C,~~ \sprod{e_\la}{e_\ka}=\ze_{\ka\la}
=\ol{\ze_{\la\ka}}\cr
\end{eq}The dual isomorphism $W\cong W^T$, induced by a nondegenerate
product, allows Dirac's bra-ket notation
\begin{eq}{l}
\left.\begin{array}{rl}
e^\ka&=\rstate{e^\ka}\cr
e_\ka&=\lstate{e^\la}\ze_{\la\ka}\cr\end{array}\right\}\then
\hbox{dual product: }
\de_\mu^\ka=\dprod{e_\mu}{e^\ka}=\sprod{e^\la}{e^\ka}\ze_{\la\mu}
= \ze^{\ka\la}\ze_{\la\mu}\cr
\end{eq}Using the dual isomorphism,
which is antilinear for a sesquilinear form $\ze$,
a linear transformation of $W$ can be expressed in the
bra-ket notation
\begin{eq}{rl}
f:W\map W,~
f=f^\la_\ka  e^\ka\ox e_\la
&=\rstate {e^\ka}\ze_{\la\mu}f^\mu_\ka\lstate {e^\la}\cr
\lstate{e^\la}f\rstate{e^\ka}&=f^\ka_\mu\ze^{\mu\la}=f^{\la\ka}\cr
\end{eq}

Orthonormal bases $\{e^a\mid a=1,\dots,n\}$ are
related to the basis $\{e^\ka\}$
by a $W$-auto\-mor\-phism $\xi$ ($n$-bein in the Hilbert space)
\begin{eq}{l}
\sprod{e^b}{e^a}=\de^{ab},~~\left\{\begin{array}{rlrl}
\xi:& W\map W,&e^\ka=\rstate{e^\ka}&\mape \xi_a^\ka \rstate{e^a}\cr
\xi^{-1T}:& W^T\map W^T,&e_\ka&\mape (\xi^{-1})_\ka^a e_a\cr
&&e_\ka\ze^{\ka\la}=\lstate{e^\la}&\mape (\xi^\star)_a^\la \lstate{e^a}\cr\end{array}
\right.
\end{eq}Bases of translation eigenstates
describing  unstable particles have not to be orthonormal.
An unstable particle collective $W\lrmap\cl P_W$ comes with its Hilbert-bein
$\xi_W$.

The state space metric
is the absolute square of the Hilbert-bein
\begin{eq}{rlrl}
\sprod{e^\la}{e^\ka}&= \ze^{\ka\la}=(\xi^\star)_b^\la\de^{ab}\xi_a^\ka,&
\ze&=\xi\de\xi^\star\cr
\sprod{e_\la}{e_\ka}&= \ze_{\ka\la}=(\xi^{-1})_\la^b\de_{ab}
(\xi^{-1\star})^a_\ka,&
\ze^{-1}&=\hat\xi\de\hat\xi^\star,~~\hat\xi=\xi^{-1\star}
\end{eq}The positivity of the Hilbert product is
seen in the matrix product form - any product $ff^\star$ of $(n\x n)$-matrices
is positive, i.e. has positive spectrum.

The Hilbert-bein $\xi$ arises in inner automorphisms for  linear
transformations - in the bra-ket formulation
\begin{eq}{l}
\lstate{e^\la}f\rstate{e^\ka}
=(\xi^\star)^\la_b\lstate{e^b}f\rstate{e^a}\xi^\ka_a
\end{eq}

Obviously, the structures above with the bra-ket formalism
and the transition from orthonormal to general bases
constitute the sesquilinear product analogue of the more
familiar structures with a bilinear real metric $g$,
e.g. for real 4-dimensional spacetime the raising and
 lowering of indices with $g$ and $g^{-1}$. The metric is the
 square of the diagonalizing
tetrad (vierbein) $h$
\begin{eq}{l}
g^{\mu\nu}=h_j^\nu\eta^{kj}h_k^\mu,~~g=h\eta h^T
\end{eq}with flat Minkowski space orthogonal  metrical matrix
$\eta={\scriptsize\pmatrix{1&0\cr0&-\bl1_3\cr}}$.
The transposition $T$
in the real bilinear case is replaced by the conjugate transposition
$\star$ for the complex Hilbert space.
The spacetime metric  discriminant
$\det g=-(\det h)^2$ has its analogue in the discriminant
of the Hilbert product
\begin{eq}{l}
\det \ze=|\det\xi|^2
\end{eq}which is used for the probability normalization of
particle transition amplitudes.

What is not analogue for spacetime metric and vierbein, on the one
side, and Hilbert space product and Hilbert-bein, on the other side,
is the real 4-dimensional spacetime dependence of the tetrad (metric)
$x\mape h(x)$
which has no counterpart in the Hilbert-bein and state space metric.
In addition there is the important
difference that the bilinear metrical matrix
 represents  a tensor $g(x)\in \M(x)\ox\M(x)$
of the tangent Minkowski translations $\M(x)\cong\R^4$ whereas
the sesquilinear scalar product matrix  of the Hilbert space
is no tensor  $\ze\notin W\ox W$.
Raising and lowering indices with the Hilbert space metric $\ze$, e.g. in
$\lstate{e^\la}f\rstate{e^\ka}=f^\ka_\mu\ze^{\mu\la}$,
changes bilinearity into sesquilinearity.

A spacetime tetrad  represents (at each spacetime point $x$)
the classes of the
real 10-dimensional symmetric space with the Lorentz groups in the general
linear group
\begin{eq}{l}
\GL(\R^4)/\O(1,3)\cong \D(1)\x\SL_0(\R^4)/\SO_0(1,3)
\end{eq}with the `overall' dilatation group $\D(1)=\exp\R$
and the orthochronous Lorentz
group $\SO_0(1,3)$. The classes are characterized
by the value of the similarity invariants which can be found in the
coefficients of the characteristic polynomial
$\det(\log h-X\bl1_4)$ for the tetrad generator.
The tetrad manifold
has two
continuous invariants $\{\De_0,\De\}$
which can be obtained also
by diagonalization of the symmetric metrical matrix
$g=g^T$ with an orthogonal transformation $O$ and
and a double hyperbolic dilatation transformation $D$, the
latter one equalizing the dilatations for the three space directions
\begin{eq}{rl}
g&=O~ D~\diag g
 ~D^T O^T\hbox{ with }
 O\in\SO(4),~D\in\SO_0(1,1)^2\cr
\diag g&=
{\scriptsize\pmatrix{e^{2\De_0}&0\cr 0&-\bl 1_3e^{2\De}\cr}}
={\ell^2\over\sqrt{c}}
{\scriptsize\pmatrix{{1\over\sqrt{c^3}}&0\cr 0&-\bl 1_3
\sqrt{c}\cr}}
\end{eq}The  two continuous invariants for the
local rescaling of  time
and  position (local time and local length unit)
\begin{eq}{l}
  (dx_0,d\rvec x)\mape  (e^{\De_0(x)}dx_0,e^{\De(x)}d\rvec x),~~
  e^{\De_0(x)}={\ell(x)\over c(x)},~e^{\De(x)}=\ell(x)
\end{eq}arise from the  invariant
$\det h={\ell^4\over c}$ for the overall dilatation
$\D(1)$ and  the  invariant $c$ (local velocity unit)
for the subgroup $\SO_0(1,1)$ in
$\SL_0(\R^4)/\SO_0(1,3)$.

As for the Hilbert space metric,
a Hilbert-bein represents a class of the real $n^2$-dimensional
manifold with the unitary groups in the general linear group
\begin{eq}{l}
\GL(\C^n)/\U(n)\cong\D(1)\x \SL(\C^n)/\SU(n)
\end{eq}with one invariant for $\D(1)$ and $n-1$ invariants
for the special factor. All $n$-invariants
(similarity invariants of the Hilbert-bein generator
in $\det[\log \xi-X\bl1_n]$) are taken from a continuous
spectrum and can be found with the manifold isomorphy
\begin{eq}{rcrcl}
\SU(n)&\cong&
\SO(2)^{n-1}&\x&\SU(n)/\SO(2)^{n-1}\cr
\SL(\C^n)/\SU(n)&\cong&
\SO_0(1,1)^{n-1}&\x&\SU(n)/\SO(2)^{n-1}\cr
\end{eq}by a special unitary diagonalization
of the hermitian scalar product matrix
$\ze=\ze^\star$
\begin{eq}{rl}
\ze&=\xi\de\xi^\star=U~\diag\ze~
U^\star\hbox{ with }
U\in\SU(n)\cr
\diag\ze&= e^{2\De_0}{\scriptsize\pmatrix{
e^{2\De_1}&\dots &0\cr
&\dots&\cr
0&\dots&e^{2\De_n}\cr}}\in\D(1)\x\SO_0(1,1)^{n-1},~~{\SUM_{k=1}^n}\De_k=0
\end{eq}The non-orthogonality of unstable particles gives rise
to invariants $\{\De_k\}_{k=1}^n$
in a normalized Hilbert-bein $\xi$,
$\det\xi=1$, which are  related to the characteristic decay parameters.

\section{Unstable States and Particles (Part 2)}

\subsection{The Unitarity of Particles}

For stable particles with mass $m$,  Wigner's definition \cite{WIG} is used,
characterizing a particle as a vector acted upon with a
unitary irreducible representation of the Poincar\'e group
which - because of the noncompact nonabelian degrees of freedom -
has to be infinite dimensional as expressed by the
integral above $\int_{\R^3} {d^3p\over(2\pi)^3p_0}$
where $\R^3$ for the momenta parametrizes the boost cosets $\SO_0(1,3)/\SO(3)$.
With Wigner's definition, confined quarks are no particles,
they are no eigenvectors
with respect to the spacetime translations, i.e. they have no
invariant translation eigenvalue (particle mass).

For masses $m^2\ge 0$,   the representations of the
 Poincar\'e covering group $\SL(\C^2)\sx\R^4$
with the orthochronous Lorentz covering group
$\SL(\C^2)/\{\pm\bl1_2\}\cong \SO_0(1,3)$
are induced from the unitary representations of the
direct product little
groups with the spin or polarization group for the space rotations
\begin{eq}{rl}
\hbox{for }m^2>0:&\SU(2)\x \R\map\U(1+2J)\cr
\hbox{for }m^2=0:&\SO(2)\x \R\map\U(1+2|J_3|)
\end{eq}The translations in the direct product groups  can be taken
to be  time translations, e.g. in a rest system for $m^2>0$.
They come in harmonic oscillator
representations as given in the 1st section
and represented in the phase group, i.e.
in $\U(1)\cong\U(1+2J)/\SU(1+2J)$
\begin{eq}{l}
\R\ni t\mape e^{iE t}\in\U(1)\hbox{ with $E\in\R$}
\end{eq}

In contrast to the compact position rotations
the noncompact translations have also representations in noncompact groups.
Unstable particles  with a nontrivial positive width are orbits
of not unitary irreducible time representations
 \begin{eq}{l}
\R_+\ni t\mape D(t)=e^{(iE-{\Ga\over 2})t}\notin\U(1)\hbox{ with $\Ga>0$}
\end{eq}They can be used only for the future monoid $\R_+$.
Unitarity as necessary for a complex representation of a real group
 is restored
by taking the direct sum with the anti-representation
(inverse-conjugated) for the past monoid $\R_-$
 \begin{eq}{l}
\R_-\ni t\mape  D(-t)^\star=e^{(iE+{\Ga\over 2})t}\notin\U(1)
\end{eq}The resulting
representation is  indefinite unitary \cite{S012}
\begin{eq}{l}
\R\ni t\mape D(t)\pl D(-t)^\star =
e^{iEt}{\scriptsize\pmatrix{
e^{-{\Ga\over 2}t}&0\cr
0&e^{+{\Ga\over 2}t}\cr}}
\in\U(1,1)\subnoteq\GL(\C^2)
\end{eq}leaving invariant the indefinite
square in  the off-diagonal isotropic basis
\begin{eq}{l}
\ze_{(1,1)}: \C^2\x\C^2\map\C,~~\ze_{(1,1)}\cong
{\scriptsize\pmatrix{0&1\cr 1&0\cr}}
\end{eq}which cannot be used for a Hilbert space product. It is possible to use
a 2nd decomposable conjugation $\U(1)\x \U(1)\subnoteq\U(2)$
with scalar product $\ze_2\cong {\scriptsize\pmatrix{1&0\cr 0&1\cr}}$
in the 2-dimensional  space.
Then unstable particles are described by non-unitary
future monoid representations
in a Hilbert space, i.e. - in contrast to stable particles -
the scalar product invariance group defining the probability
does not contain the represented time translations. Probabilities have a
nontrivial time dependence.

Such a $\U(1,1)$-representation of the translations can be used - as for
stable particles - to define corresponding representations of the
 Poincar\'e  monoid $\SL(\C^2)\sx\R^4_+$
with  $\R^4_+$ the future  spacetime translation cone.
In addition to the energy width there arises also a momentum spread
and, therewith, nontrivial spin mixtures for unstable particles \cite{BS01}.

These were some short remarks to bypass a not satifactorily
solved problem - how
to reconcile
the  different unitarities for rotations (definite) and
spacetime translations (possibly indefinite)
with the  probability
interpretation (necessarily definite).

\subsection{Nondecomposable Particle Collectives}

In the `huge' Hilbert space with all particles
there are
- neglecting the momentum dependence $\rvec p\in\R^3$ -
1-dimensional subspaces connected with
stable  particles and higher dimensional ones for
decay collectives. With respect to probabilities and expectation values, those
subspaces have to be
considered as a `whole'.
This can be seen in some analogy to a relativistic spacetime
vector with
many bases for time and space projections $x=(x_0,\rvec x)$,
but with only one Lorentz length $x^2$.

It is assumed that the `huge' Hilbert space has a basis with
particle states (translation eigenstates), stable
and unstable.
It is decomposable into nondecomposable orthogonal subspaces
$\sprod{W}{U}=\{0\}$ for $W\ne U$, assumed to be finite dimensional
(always neglecting the continuous momentum dependence $\rvec p\in\R^3$).

A basis with $n$ particle states in an
orthogonally nondecomposable subspace $W\cong\C^n$
with a corresponding positive  scalar product matrix
$\ze_W=\xi_W\xi_W^\star $
and Hilbert-bein $\xi_W$ will be probability-normalized by its discriminant, $\det \ze_W=1$,
and with trivial phase, i.e. the Hilbert-bein  $\xi_W$ involves
maximally $(n^2-1)$-real parameters
of a noncompact class $\SL(\C^n)/\SU(n)$ with $(n-1)$ continuous invariants
 \begin{eq}{l}
\hbox{scalar product }\ze_W=\xi_W \xi_W ^\star\succeq0\hbox{ with }\det \xi_W =1
\end{eq}

A nondecomposable 1-dimensional space is the ray  of a stable particle state
\begin{eq}{l}
W=\C\rstate m
\hbox{ with }\det\ze_1=\sprod mm=1
\end{eq}An orthogonally nondecomposable space with $n\ge2$
describes a decay collective
$W=\rstate M\pl\rstate m\cong\C^n$ as discussed above:
The  Hilbert-bein as transformation
from orthogonal to particle basis can be brought to a triangular form
$\xi_W ={\scriptsize\pmatrix{\xi_d&w\cr0&\bl1_s\cr}}
$, $w\ne0$,
with a unit submatrix leaving invariant
the subspace $\rstate m\cong\C^s $
with $s\ge1$ stable states and a non-orthogonal complement
$\rstate M\cong\C^d $ with  $d\ge1$ unstable states.
The  discriminant normalization
of the full scalar product matrix
coincides with the  discriminant normalization
for the unstable states
\begin{eq}{rl}
\det \ze_W&=\sprod{\det \xi_W }{\det \xi_W }=
\sprod{M}{M}\sprod{ m}{m}-\sprod{ M}{m}\sprod{ m}{M}\cr
&=\det {\scriptsize\pmatrix{
\ze_d +ww^\star & w\cr w^\star& \bl1_s\cr}}=\det \ze_d=\sprod{\det \xi_d}
{\det \xi_d}=1
\end{eq}The projector
for the decay collective involves the inverse scalar product matrix $\ze_W^{-1}$
\begin{eq}{l}
\cl P_W=
\rstate M \ze_d^{-1}\lstate M
-\rstate M \ze_d^{-1} w\lstate m
-\rstate m w^\star \ze_d^{-1}\lstate M
+\rstate m (\bl1_s+w^\star \ze_d^{-1} w)\lstate m\cr
\end{eq}The projectors for the non-orthogonal subspaces are
\begin{eq}{l}
\cl P_{\rstate M}=\rstate M (\ze_d+ww^\star)^{-1}\lstate M,~~
\cl P_{\rstate m}=\rstate m\bl1_s\lstate m
\end{eq}

The probability  normalization
  for the kaon system
with the discriminant is collective: It involves the
decay parameter, i.e. the non-orthogonality $\de$
with $0<\de^2<1$
\begin{eq}{rl}
\det \ze_2&
=\sprod{M_S  }{M_S  }\sprod{M_L  }{M_L  }
-\sprod{M_S  }{M_L  }\sprod{M_L  }{M_S  }\cr
&=\det{1\over |N|^2}
{\scriptsize\pmatrix{1&\de\cr
\de &1\cr}}={1-\de^2\over |N|^4}=1\then |N|^2=\sqrt{1-\de^2}
\end{eq}and differs
 from the individual probability normalization for each particle
which would be given by $|N|^2=1$.
The continuous invariant $\De$ related to the rank 1 of the
Hilbert-bein manifold $\SL(\C^2)/\SU(2)\cong \SO_0(1,1)\x\SU(2)/\SO(2)$ is seen in the $\SO_0(1,1)$-adapted
parametrization of the kaon-scalar product
\begin{eq}{l}
{\scriptsize\pmatrix{
\sprod{M_S  }{M_S  }&\sprod{M_S  }{M_L  }\cr
\sprod{M_L  }{M_S  }&\sprod{M_L  }{M_L  }\cr}}
={\scriptsize\pmatrix{
\cosh\De&\sinh\De\cr
\sinh\De&\cosh\De\cr}}\sim
{\scriptsize\pmatrix{
e^\De&0\cr
0&e^{-\De}\cr}}, ~~\tanh \De =\de
\end{eq}

Neglecting  weak and electromagnetic
interactions, the neutron and pion, e.g., are stable.
Taking into account
the mentioned  interactions,  each of these particles
constitutes the decaying subspace of a
decay collective, e.g. for the
neutron $\{\rstate n,\rstate{p,e,\ol \nu_e}\}$
with a nontrivial projection $\sprod n{p,e,\ol \nu_e}$.
`Switching on' all interactions there seems to exist only a small
number of high dimensional orthogonally nondecomposable particle collectives which
span the particle Hilbert space. Their
particle representatives with lowest mass
(the lowest step in the staircase)
 reflect the few invariants of relativistic
particle physics, i.e.
the  mass for spacetime translations $\R^4$,
the rotation invariants, characterizing spin
 $\SU(2)$ for massive particles
and polarization $\SO(2)$ for massless ones
and the electromagnetic charge number for
 a phase group $\U(1)$

{\scriptsize
\begin{eq}{c}
\begin{array}{|c||c|c|c|c|c|}\hline
\hbox{particle}
&\hbox{mass for}&\hbox{spin }\SU(2)\hbox{ or}&\hbox{charge}
&\hbox{fermion number}\cr
\hbox{(lowest mass}
&\hbox{translations }\R^4&\hbox{polarization }\SO(2)&\U(1)&\U(1)\cr
\hbox{representative)}&m^2&J\hbox{ or }\pm |J_3|&Q&F\cr\hline\hline
\hbox{photon}&0&\pm 1&0&0\cr\hline
\hbox{(anti)proton}&>0&{1\over2}&\pm 1&\pm 1\cr\hline
\hbox{(anti)electron}&>0&{1\over2}&\mp 1&\pm 1\cr\hline
\hbox{(anti)neutrino}&>0\hbox{ or }=0~(?)&{1\over2}
\hbox{ or }\pm {1\over 2} (?)&0&\pm 1\cr\hline
\end{array}\cr\cr
\hbox{\bf invariants for nondecomposable collectives}

\end{eq}
}\noindent If the proton is stable, there has to be  an
additional invariant,
usually related  to fermion number $F$ conservation
which is taken care of with the different
 association of charge and fermion
number for proton with  $Q+F=2$ and electron with $Q+F=0$.
In addition, there may exist invariants for
the leptonic phases - electronic, muonic, tauonic.

Stable and unstable particle states come on the same level
as Hilbert space directions - stable
particles, e.g. the  electron, are  not `more fundamental' than
unstable ones, e.g. the muon or the pion.
An $S$-matrix with only stable
in- and out particle
state vectors where  the unstable ones are taken
care of as intermediate fictive poles
only \cite{WEIN1} is against a democratic treatment.
Depending of the degree of approximation to the distinction of
 stable-unstable as quantified in
  the magnitude of the off-diagonal entries $w$ in the
 Hilbert-bein  $\xi_W $
one may work with a larger or a smaller number of
nondecomposable collectives,
 i.e. with  a smaller or a larger number of stable particles.

\subsection{(Non)Unitary $S$-Matrix for Unstable Particles}

As an example, how the collective higher dimensional
structure affects the
probability interpretation,
 the unitarity structure of the $S$-matrix, involving
a scattering with  unstable particles, is considered.

Starting from a free Hamiltonian $H_0$ acting on a Hilbert space
with an eigenvector basis
\begin{eq}{l}
H_0\rstate E=E\rstate E
\end{eq}the in and out states for an interaction Hamiltonian $H$
are assumed to be constructable by inner automorphisms with the
Moeller operators $\Om_\pm$ for infinite future and past \cite{WEIN1}
\begin{eq}{l}
H=\Om_\pm H_0\Om_\pm^{-1}\then
H\rstate{ E_\pm}=E\rstate {E_\pm}\hbox{ for }\rstate {E_\pm}
=\Om_\pm\rstate {E_\pm}
\end{eq}The scattering operator
is the product of the Moeller operators
\begin{eq}{l}
S=\Om_+^\star\Om_-
\end{eq}

In quantum mechanics the Moeller operators are assumed to
arise as limits
\begin{eq}{l}
\Om(t)=
e^{iHt}e^{-iH_0t},~~
\Om_\pm=\lim_{t\to\pm\infty}\Om(t),~~
\end{eq}Unitarity for  hermitian
Hamiltonians is assumed to survive the limit
\begin{eq}{l}
\hbox{if }H_0=H_0^\star, ~H=H^\star\then
\Om(t)^\star=\Om(t)^{-1},~~
\Om_\pm ^\star=\Om_\pm^{-1}\then S^\star=S^{-1}
\end{eq}

The scattering amplitudes
are not matrix elements of  linear transformations, but scalar products
 of in and out states - sesquilinear, not bilinear structures. They
start with the scalar product matrix
of the free particles
 \begin{eq}{rl}
\sprod{E^\la_+}{E^\ka_-}
=\lstate{E^\la}S\rstate {E^\ka}=\sprod{E^\la}{E^\ka}-2i\pi
\lstate{E^\la}T\rstate {E^\ka},~~S=\bl 1-2i\pi T
\end{eq}If a decay collective is involved,
the $S$-matrix $S^{\ka\la}$ is not unitary
\begin{eq}{l}
S^{\ka\la}=\lstate{E^\la}S\rstate{ E^\ka}=
\ze^{\ka\la}+\dots =\xi^\la_b\de^{ba}(\xi^\star)^\ka_a+\dots
\end{eq}Unitarity is expected for the  $S$-matrix
$S^a_c$ transformed with the Hilbert-bein
from  a non-orthonormal particle basis
into an orthonormal non-particle basis
\begin{eq}{l}
\hat\xi_\la^b \lstate{E^\la}S\rstate{ E^\ka}(\hat\xi^\star)_\ka^a=
\lstate{b}S\rstate{a}=S^a_c\de^{cb}\cr
\end{eq}

\centerline{\bf Acknowledgment}
\vskip2mm
I am indebted to Walter Blum for discussions
on the `identity overlap' for decay collectives as seen in the
nontrivial scalar products involving
unstable particles and their decay products.

\end{document}